\documentclass[manuscript,screen]{acmart}
\AtBeginDocument{%
  \providecommand\BibTeX{{%
    \normalfont B\kern-0.5em{\scshape i\kern-0.25em b}\kern-0.8em\TeX}}}

\setcopyright{acmlicensed}
\copyrightyear{2018}
\acmYear{2018}
\acmDOI{XXXXXXX.XXXXXXX}

\acmConference[Conference acronym 'XX]{Make sure to enter the correct
  conference title from your rights confirmation emai}{June 03--05,
  2018}{Woodstock, NY}
\acmBooktitle{Woodstock '18: ACM Symposium on Neural Gaze Detection,
 June 03--05, 2018, Woodstock, NY} 
\acmISBN{978-1-4503-XXXX-X/18/06}

\begin{document}
\pdfoutput=1

\title[Display in the Air]{Display in the Air: Balancing Distraction and Workload in AR Glasses Interfaces for Driving Navigation}

\author{Xiangyang He}
\email{xiangyanghe@link.cuhk.edu.cn}
\orcid{0009-0006-9055-2547}
\affiliation{%
  \institution{The Chinese Univerisity of Hong Kong (Shenzhen)}
  \streetaddress{No 2001 Longxiang Boulevard}
  \city{Shenzhen}
  \state{Guangdong Province}
  \country{China}
  \postcode{43017-6221}
}

\author{Keyuan Zhou}
\email{zhoukeyuan@oppo.com}
\orcid{0000-0003-3484-8897}
\affiliation{%
  \institution{OPPO Research Institute}
  \streetaddress{GWJ2+RH4}
  \city{Shenzhen}
  \state{Guangdong Province}
  \country{China}}

\renewcommand{\shortauthors}{He, et al.}

    \begin{abstract}
    Augmented Reality (AR) navigation via Head-Mounted Displays (HMDs), particularly AR glasses, is revolutionizing the driving experience by integrating real-time routing information into the driver's field of view. Despite the potential of AR glasses, the question of how to display navigation information on the interface of these devices remains a valuable yet relatively unexplored research area. This study employs a mixed-method approach involving 32 participants, combining qualitative feedback from semi-structured interviews with quantitative data from usability questionnaires in both simulated and real-world scenarios. Highlighting the necessity of real-world testing, the research evaluates the impact of five icon placements on the efficiency and effectiveness of information perception in both environments. The experiment results indicate a preference for non-central icon placements, especially bottom-center in real world, which mostly balances distraction and workload for the driver. Moreover, these findings contribute to the formulation of four specific design implications for augmented reality interfaces and systems. This research advances the understanding of AR glasses in driving assistance and sets the stage for further developments in this emerging technology field.
    \end{abstract}



\begin{teaserfigure}
  \includegraphics[width=\textwidth]{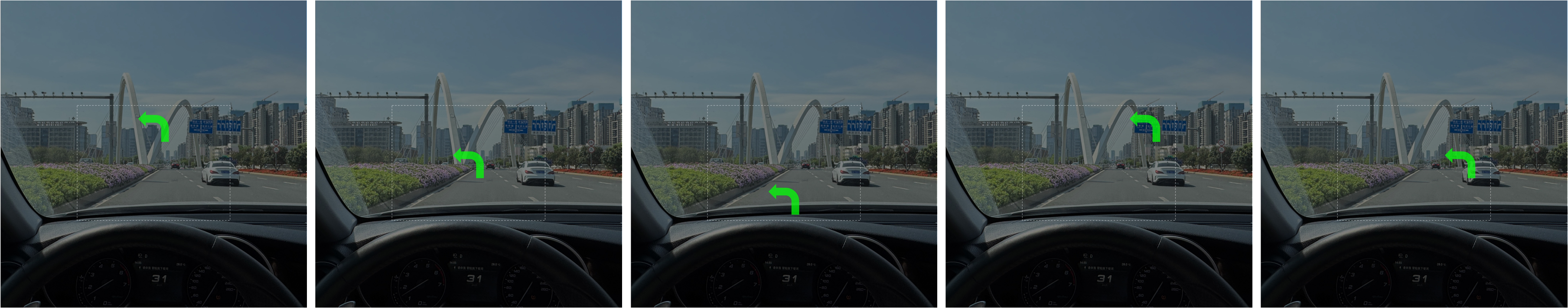}
  \caption{Navigation Icon Position Examples in AR Glasses Assisted Driving.}
  \Description{Using AR Glasses to Access Navigation Information.}
  \label{fig:teaser}
\end{teaserfigure}

\received{20 February 2007}
\received[revised]{12 March 2009}
\received[accepted]{5 June 2009}

\maketitle

\section{Introduction}

Augmented Reality (AR) is recognized for seamlessly integrating virtual content with real-world scenes \cite{mekni2014augmented}, a capability that has broad applications across various contexts and hardware technologies \cite{gabbard2019ar}. Particularly, in the realm of automotive human-computer interaction (HCI), AR is gaining popularity not just for enhancing road safety \cite{kim2013exploring,10.1145/3543174.3546089} and navigational aids \cite{topliss2018establishing, 10.1145/2799250.2799253}, but also for improving driving and passenger experiences \cite{riegler2021augmented}. These implementations have ranged from devices that combine camera views with display interfaces \cite{shahriar2018camera} to see-through technologies like Head-Up Displays (HUDs) \cite{10.1145/3349263.3351911,10.1145/3581961.3609874}. However, given the constraints of HUDs like size, there is increasing interest in Head-Mounted Displays (HMDs), including portable AR glasses. These developments are poised to drive further innovations in the automotive sector \cite{10.1145/3581961.3609870}. 

The majority of research on HMDs has focused on platforms like the Microsoft HoloLens \cite{kun2017calling,10.1145/3581961.3609870,9020354}. However, there remains a significant research gap concerning the integration of lightweight, portable AR glasses within automotive environments. Compared to HoloLens, which weighs over 500 grams, portable AR glasses are significantly lighter, typically weighing around 100 grams or less \cite{du2023comfort}. The excessive weight of heavy HMDs can lead to musculoskeletal discomfort, potentially leading to long-term consequences \cite{chen2021human, astrologo2024determining}. These issues are particularly concerning for drivers who need to continuously scan their environment to ensure road safety. Moreover, compared to the 500 nits brightness offered by HoloLens, some new portable AR glasses products can reach a brightness level of over 1000 nits by Micro-LED \cite{vogel2009hypoled,yu2022gallium}. This superior brightness, combined with their lighter weight, makes portable AR glasses more practical for real-world navigation applications and research.

Driving navigation exemplifies a dual-task scenario \cite{imamov2020display}, where AR can significantly enhance efficiency by allowing drivers to acquire information swiftly while maintaining focus on the road, thereby requiring minimal cognitive effort \cite{bell2002information,haeuslschmid2016design}. Unlike simpler tasks such as checking the date or weather, navigation demands sustained attention. It is crucial for drivers to prioritize road safety and avoid becoming overly absorbed in AR interactions. Thus, AR interfaces must be designed to be easily noticeable without being distracting \cite{gabbard2014behind}. 

This study progresses by investigating AR navigation interfaces that could minimize both distraction and workload. The goal is to ensure drivers can effectively utilize AR technology without compromising their alert to the surroundings. Additionally, four practical suggestions are provided for designers to make a quick inspection of their AR interfaces. The contributions of this research are outlined as follows:

\begin{itemize}
    \item This study finds a non-central placement preference in AR driving navigation scenarios, validated through both simulated and real-world settings, demonstrating an appropriate balance between distraction and workload.
    \item The experiment innovatively utilized a portable AR glasses in a real-world navigation scenario, promoting practical applications of HMDs in driving navigation and inspiring further design considerations.
    \item By integrating qualitative insights and quantitative data, this study provides a comprehensive understanding of user preferences and a framework for designers to evaluate their AR interfaces and systems.
\end{itemize}

\section{Related Work}

\subsection{AR in Driving Context}
Integrating AR into car navigation tasks significantly enhances the driving experience by reducing mental workload \cite{bauerfeind2021navigating,10.1145/2799250.2799253} and improving driving performance \cite{ma2021impact}. In the driving context, AR implementation can be facilitated through three primary mediums: AR Head-Down Displays (AR-HDDs), AR Head-Up Displays (AR-HUDs), and AR Head-Mounted Displays (AR-HMDs). AR-HDDs, considered the most accessible approach, are typically achieved through camera-view AR, assisting drivers in tasks such as navigation and ADAS. Subsequently, AR-HUDs emerged as a superior assistive tool. Nwakacha et al. \cite{nwakacha2013evaluating} discover that HUDs are considered less physically and mentally demanding compared to HDDs. Horrey et al.  \cite{horrey2003effects} and Medenica et al. \cite{medenica2011augmented} report that HDDs decreased driving performance in comparison to HUDs. Consequently, industrial manufacturers and research laboratories have endeavored to achieve HUD functionality primarily through two methods: windscreen projectors and comprehensive AR windscreen integration \cite{10.1145/3581961.3609870}, albeit with technical challenges \cite{gabbard2014behind}. Moreover, the limited Field-of-View (FOV) of AR-HUDs, covering only the front lane view of the driver, presents challenges in providing warnings for objects at broader angles around the driver \cite{10.1145/3581961.3609870}. AR-HMDs show more potential to solve the above problems and make drivers comprehend information more quickly and easily than conventional HUDs \cite{bauerfeind2021navigating}. With the rapid growth of AR-HMDs industry, transitioning from laboratory prototypes like Google Glass \cite{sawyer2014google} to commercial products like HoloLens series by Microsoft \footnote{\href{https://learn.microsoft.com/en-us/hololens/hololens1-hardware}{HoloLens (1st gen) hardware}}\footnote{\href{https://www.microsoft.com/en-us/hololens}{HoloLens 2 - Mixed Reality Technology for Business}}, and more recent entries such as RayNeo X2 \footnote{\href{https://www.rayneo.com/products/tcl-rayneo-x2}{TCL RayNeo - next generation AR and XR Glasses}} and OPPO Air Glass series \footnote{\href{https://www.oppo.com/en/newsroom/press/oppo-air-glass/}{OPPO Introduces Air Glass, Featuring Creative Cicada Wing Design and Self-designed Spark Micro Projector}}\footnote{\href{https://communityin.oppo.com/thread/1222179266759491585}{Introducing Smart Glass | AIR GLASS 2}}\footnote{\href{https://www.oppo.com/en/newsroom/press/oppo-unveils-new-oppo-air-glass-3/}{OPPO unveils new OPPO Air Glass 3 at MWC 2024}}, demonstrating significant technological evolution. These products, resembling standard glasses and equipped with built-in navigation applications, represent a more practical approach for in-car usage. For instance, the RayNeo X2 employs simultaneous localization and mapping (SLAM) in its GPS navigation system \cite{kemeny2023virtual}, yielding a positive user experience.

\subsection{Navigation Information Placement}

Given that AR interfaces can either enhance or detract from the user experience based on the design of graphical elements \cite{kim2022assessing}, it is crucial to mitigate distraction by minimizing information overload and avoiding `clutter'. This necessitates a deeper understanding of how the size, shape, color, and placement of cues can promote quicker recognition and reduce occlusion \cite{10.1145/2799250.2799253,kim2013exploring}.

Our research primarily addresses cue placement. Rzayev et al. \cite{rzayev2018reading} exam positioning in top-right, center, and bottom-center locations by reading text, finding that top-right placement heightens subjective workload and impaired comprehension, with a preference for bottom-center and center locations. Imamov et al. \cite{imamov2020display} assess the impact of glanceable interface positions on comfort and task-switching time, concluding that placements at or below eye level enhance speed and comfort. Lee and Woo \cite{lee2022exploring} investigate AR notification placements in interactions with virtual avatars, noting that top-left positions should be avoided due to increased response times and task load, whereas middle-right position is more effective, leading to faster responses and reduced task load. Chua et al. \cite{chua2016positioning} explore display positions on optical see-through HMDs in dual-task scenarios, finding that notifications are more rapidly noticed at middle and bottom center positions. Yet, top and peripheral positions are deemed more comfortable and less intrusive, with the middle-right position offering an optimal balance between performance and comfort. To circumvent user fatigue from assessing nine different locations, we focus on promising locations identified in prior research: \textit{top-center, middle-center, bottom-center, top-right, and middle-right}, and incorporate these into the AR glasses platform, which currently lacks extensive research in this area.

\subsection{Simulated and Real-World AR Driving Studies}

In AR driving studies, researchers often utilize driving simulators, with monitor-based setups \cite{alves2013forward,chen2017eliminating,fremont2020adaptive} or VR systems \cite{10.1145/3321335.3324947,charissis2014enhancing}, while some investigations employ in-vehicle setups \cite{10.1145/2799250.2799253,bram2020collision,calvi2020effectiveness}. However, these methods face validity concerns and might omit some important factors related to the interface design. Detjen et al. \cite{10.1145/3543174.3546089} report that the ecological validity of simulated driving leads to low perceived realism and medium involvement among participants. Additionally, real-world scenarios exhibit greater complexity and variability. Given the unpredicted risks associated with testing a prototype in real driving conditions, employing a co-pilot navigation task serves as a viable alternative method. Drawing inspiration from Bolder et al. \cite{bolder2018comparison}, our experiments encompass both simulated and real driving scenarios, utilizing AR glasses as the navigation platform to explore the commonalities and differences between the two settings.

\section{Study}

This research are divided into two studies: \textbf{Pilot Study} evaluates usability concerns with the current navigation interface design in the real-world navigation scenario. Based on the findings, \textbf{Formal Study} adopted a methodology to explore the ideal positioning of navigation information on AR glasses in both simulated and real driving scenarios. 

\subsection{Pilot Study and Preliminary Findings}

This part was conducted in a real-world setting, where 10 participants sat on the co-pilot seat of a moving vehicle and reviewed the AR navigation interface. One example is depicted on the left of Figure \ref{fig:icon collection}. Following a 10-minute journey, participants were requested to rate their overall satisfaction using a 7-point scale and provide reasons for their ratings.

The analysis revealed that the mean overall satisfaction was 3.20, with a standard deviation of 1.62. Common insights, identified by more than half of the participants, were derived from their feedback. Although the majority acknowledged that the prominence of the navigation information facilitated adaptation to the interface, 9 participants reported distractions caused by the centralized placement of this information. This observation led to further investigations into the optimal information placement of the navigation interface, emphasizing the balance between visual prominence and minimal distraction. Additionally, 7 participants noted that the volume of information presented was overwhelming during the journey. They suggested that textual information was less critical for real-time navigation than directional arrows; consequently, only graphical information was incorporated into the formal study. To further reduce workload, 4 of them recommended that textual information should either be displayed for shorter periods or appear only when the driver was not engaged in complex driving tasks. Finally, 6 participants reported disturbances caused by the interface moving synchronously with head movements, particularly during rapid head turns to scan the environment.

\subsection{Formal Experiment Design}

\begin{figure}[htbp]
  \centering
  \includegraphics[width=0.7\linewidth]{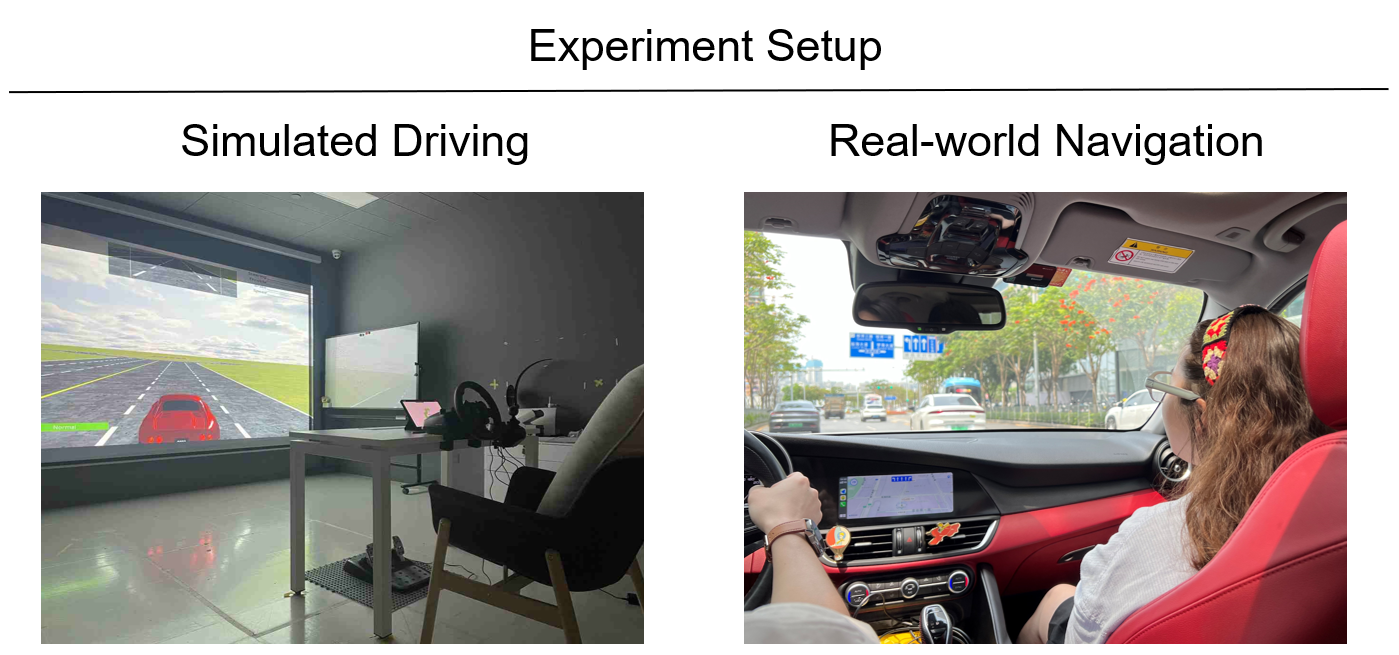}
  \caption{The simulated driving environment with a driving simulator and a projection screen (left) and the real-world navigation setup in a vehicle with AR glasses in use (right).}
  \Description{On the left, a simulated driving setup is presented, featuring a steering wheel and a projected virtual road. On the right, an individual equipped with AR glasses navigates a real vehicle, with navigational directions displayed on the AR interface.}
  \label{fig:setup}
\end{figure}

\begin{figure}[htbp]
  \centering
  \includegraphics[width=1\linewidth]{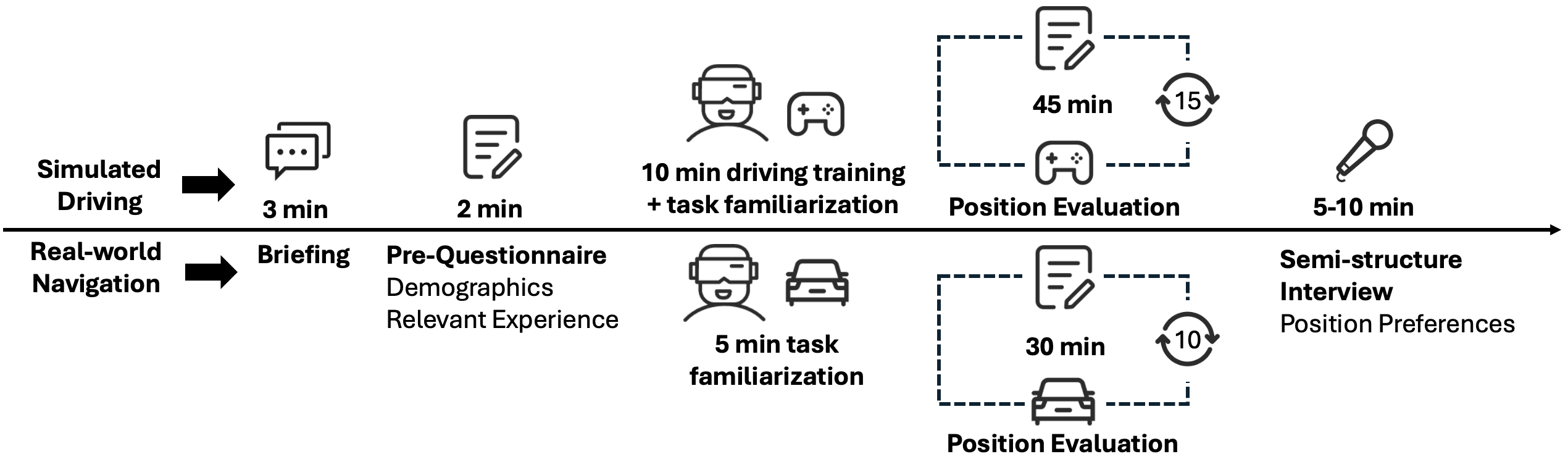}
  \caption{Procedure of the simulated driving and real-world navigation experiment.}
  \Description{The flowchart represents the experiment procedure. It starts with separate briefings for simulated driving and real-world navigation, followed by a pre-questionnaire. Participants for each experiment then undergo driving training and task familiarization. After completing the tasks, position evaluations are conducted, which are then followed by a semi-structured interview to discuss position preferences. The durations of each step are specified.}
  \label{fig:procedure}
\end{figure}

The study employs a within-subjects design across two independent scenarios, as illustrated in Figure \ref{fig:setup}. The independent variable is the positioning of the navigation icon on the AR glasses interface, distributed across five distinct locations. The dependent variables are from a \textit{Fast Position Evaluation Questionnaire (FPEQ)}, which encompasses three items rated on a 7-point scale: (1) Overall Appropriateness: the placement of the navigation icon in this position is appropriate; (2)Distraction Level: the navigation icon in this position distracts from driving/navigating tasks; (3) Workload Level: the navigation icon in this position demands mental and physical effort to be noticed.

The procedures for formal experiments are shown in Figure \ref{fig:procedure}. In the simulated driving experiment, participants engage in two preliminary driving sessions before the commencement of the formal study. The initial session aims to acclimate them to the simulator, while the subsequent session focuses on familiarizing them with the experimental tasks. Each of these preparatory sessions lasted approximately 5 minutes. During the main experiment, participants encounter five distinct positions of the navigation icon. Each icon position undergoes three evaluations in random sequences which is balanced by employing a Latin square design, totaling fifteen scenarios. Following each navigation cue and the corresponding driving task completion, participants provide verbal assessments based on the FPEQ. At the conclusion of the experiment, a semi-structured interview session was conducted to address usability concerns, during which participants were asked to indicate their preferred positions and explain their choices.

In the real-world driving scenario, due to safety considerations and inspiration from previous research, participants were not required to drive but instead to be a co-pilot, focusing on observing the road condition and reporting real-time directions at each intersection. During the whole journey, the participants would assume them as a real co-pilot without knowing that the driver would drive the same route as planned by researchers. Following the display of navigation icons in varying positions, participants complete the FPEQ to assess each icon. The experimental route spanned three kilometers, featuring ten intersections where each position was evaluated twice. Lastly, the same interview session was conducted in this scenario too.

\subsection{Participants}
The formal study engaged 22 participants, divided into two cohorts based on the experimental conditions: simulated driving and real-world navigation. The simulated driving cohort comprised 12 individuals (7 males and 5 females), ranging in age from 22 to 28 years, with a mean age of 25.16 years and a standard deviation of 1.67. Within this group, one participant had less than one year of driving experience, and three required optical correction. The real-world navigation cohort included 10 individuals (5 males and 5 females), aged between 22 and 35 years, with a mean age of 26.60 years and a standard deviation of 4.34. Within this group, two participants had less than one year of driving experience, and two required optical correction. 

Prior to the study, none of the participants had any experience with AR glasses. Participants who required optical correction were not allowed to wear AR glasses over their own spectacles, so they were provided with suitable contact lenses for the whole experiment. The payment for participation was set at 100 RMB (approximately 14 USD) per person. The study adhered to standard Institutional Review Board (IRB) protocols, ensuring informed consent, anonymity, and the safety of all participants.

\subsection{Materials}

For the simulated driving scenario, the apparatus encompassed a Logitech 920 driving simulator, an Epson projector (EPSON CB-FH52 with a resolution of 1920×1080), a 150-inch projection screen, a desktop computer, and the open-source PGDrive software \cite{li2020improving}. The real-world navigation trials employed a real car and one smartphones for recording purpose. Both scenarios utilized the OPPO Air Glass 2 as the navigation platform. This AR device weighs 38 grams and projects green graphics with a maximum brightness exceeding 1000 nits, suitable for daytime use. It features a 30° diagonal field of view, a screen-fixed design, and a resolution of 640×480. According to John S. Stahl \cite{stahl1999amplitude}, the comfortable rotation angle for the human eye is approximately ±25º. Furthermore, research conducted by John M. Franchak et al. \cite{franchak2021adapting} on target searching tasks indicate average human eye rotation angles of 12.9° and 13.5°. In the context of our experimental setup, for an icon positioned in the middle right, focusing at the center of the image would necessitate an eye rotation of approximately 10.12°, a value that falls within an acceptable range.

\begin{figure}[htbp]
  \centering
  \includegraphics[width=0.65\linewidth]{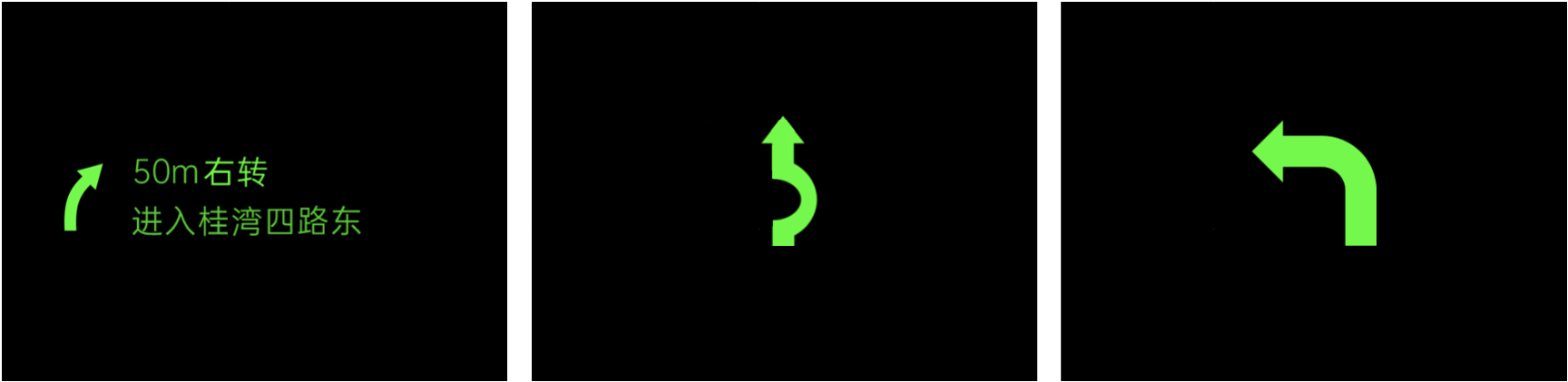}
  \caption{Examples of navigation icons used in the study: the left image displays a turn indicator with distance and directional text for interface evaluation, while the middle and right images depict simplified turn indicators for simulated and real-world scenarios.}
  \Description{Three navigation icons are presented against a transparent background. The first icon on the left includes a green arrow curving to the left with text indicating a distance of 50 meters and a prompt for a right turn. The middle icon shows a stylized dashed green arrow pointing upward, representing a straightforward direction. The right icon displays a solid green arrow curving to the left. These icons represent different of direction indicators used in the interface of navigational systems within the study.}
  \label{fig:icon collection}
\end{figure}

Figure \ref{fig:icon collection} showcases examples of the navigation interfaces for different experimental contexts. The left shows the pre-designed interfaces for evaluation in the pilot study, which is tailored for Chinese participants, featuring design language in Chinese that includes critical navigation information such as the distance to the upcoming turn and the associated road name. The middle and right show an example of icon placement at the middle-center for formal study. 

\section{Results}

The evaluation of AR glasses for driving assistance was conducted through a one-way repeated measures ANOVA, with icon position designated as a within-subjects factor in Section 4.1. In addition to the quantitative analysis, the results of semi-structured interviews were summarized to show qualitative insights regarding preferences for icon positioning in Section 4.2. 

\subsection{Information Position Analysis}

\begin{figure}[htbp]
  \centering
  \includegraphics[width=1\linewidth]{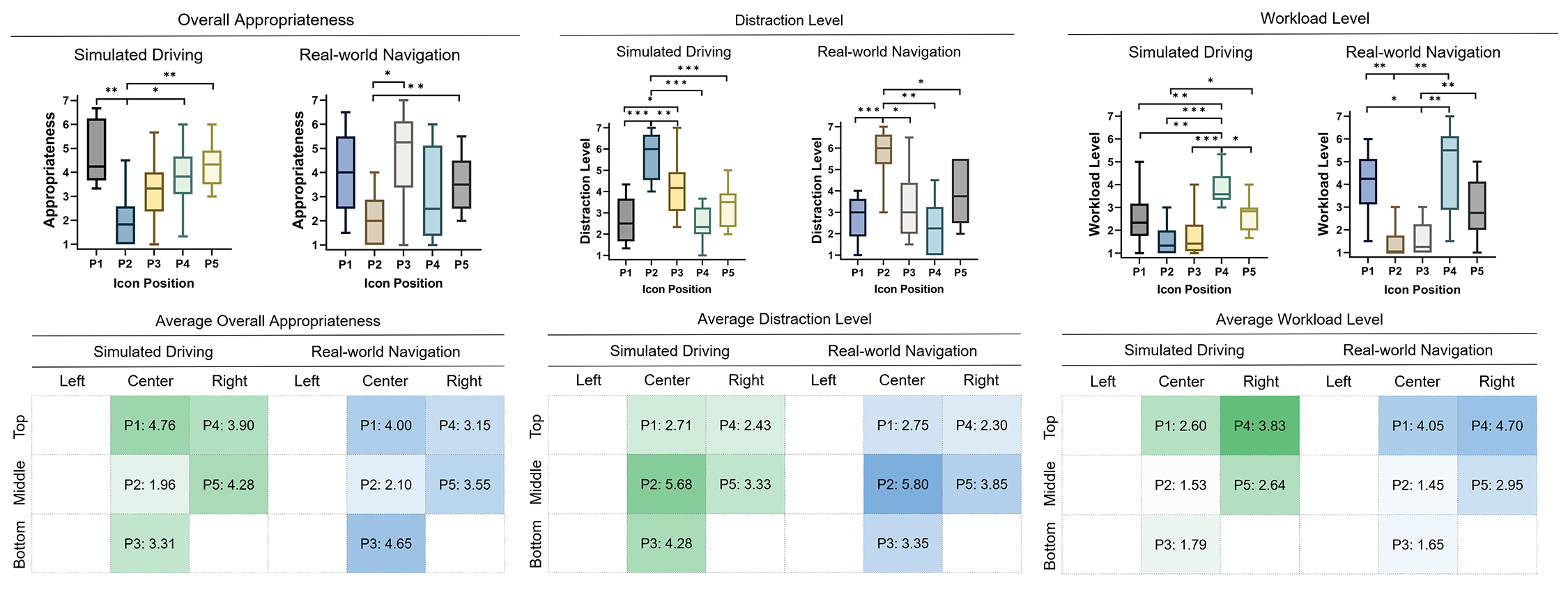}
  \caption{Box plots depicting three ratings of icon positions P1 through P5 during simulated driving (left) and real-world navigation (right). Asterisks indicate significant differences between icon positions: * p < 0.05, ** p < 0.01, *** p < 0.001. The corresponding icon position visualization comparing the average scores for icon positions in both simulated driving and real-world navigation conditions, with color intensity indicating score magnitude.}
  \Description{The figure presents a comparative analysis of icon position effectiveness across simulated driving and real-world navigation. It features six box plots, two sets of three side by side, representing Overall Appropriateness, Distraction Level, and Workload Level. Each set compares five icon positions labeled P1 to P5. Statistical significance is noted with asterisks. Beneath the box plots are six heatmaps that summarize the average ratings for icon positions, with darker colors indicating higher scores. The heatmaps are divided by scenario (simulated driving and real-world navigation) and further by icon position (top-center, middle-center, bottom-center, top-right, middle-right).}
  \label{fig:sig}
\end{figure}

Prior to the analysis, the data underwent preprocessing, where the mean of three evaluations from 12 participants in simulated driving was calculated for the simulated driving setting, and the mean of two evaluations from 10 participants in real-world navigation was calculated for the real-world navigation setting. Subsequently, the Greenhouse-Geisser correction was applied in the RM ANOVA. The results, illustrated in Tables \ref{table:combined_results}, revealed that the independent variable of location had significant main effects across three metrics in both sets of experiments. An exception was noted in the Overall Appropriateness score in the real-world navigation experiment, where the effect of location is not significant (P = 0.069). Although this result did not achieve conventional significance, its proximity to the threshold suggests a potential effect. Further analyses included multiple comparison tests for each metric, employing the Bonferroni correction. Visual representations of the data, including box plots for three ratings of icon positions and the average score visualization of icon position, are provided in Figure \ref{fig:sig}.

\begin{table}[h]
\centering
\caption{Results of Within-subjects Effects Test for Simulated Driving and Real-world Navigation}
\label{table:combined_results}
\begin{tabular}{@{}lcccccccc@{}}
\toprule
& \multicolumn{4}{c}{Simulated Driving} & \multicolumn{4}{c}{Real-world Navigation} \\
\cmidrule(lr){2-5} \cmidrule(lr){6-9}
Question & F & df & P & $\epsilon^2$ & F & df & P & $\epsilon^2$ \\ 
\midrule
Overall Appropriateness & 9.779 & 3.005, 33.056 & 0.000 & 0.471 & 3.098 & 2.017, 18.155 & 0.069 & 0.256 \\
Distraction Level & 19.826 & 2.165, 23.812 & 0.000 & 0.643 & 16.586 & 2.768, 24.913 & 0.000 & 0.648 \\
Workload Level & 20.371 & 2.776, 30.540 & 0.000 & 0.649 & 20.484 & 2.408, 21.670 & 0.000 & 0.695 \\
\bottomrule
\end{tabular}
\end{table}

\textbf{Overall Appropriateness.} This metric assesses how appropriate participants found the placements of the icon, with higher scores indicating greater satisfaction. In the simulated driving experiment, the position P2 (middle-center) is deemed significantly less appropriate than P1 (top-center, P=0.001), P4 (top-right, P=0.035), and P5 (middle-right, P=0.001), scoring an average of 1.96. For real-world navigation, P2 is also less appropriate compared to P3 (bottom-center, P=0.021) and P5 (center-right, P=0.001), with an average score of 2.10. Notably, the overall influence of location on appropriateness is marginally significant in real-world navigation (P=0.690). Therefore, the optimal positions for icon placement are \textbf{top-center (P1)} in simulated driving scenarios and \textbf{bottom-center (P3)} in real-world contexts.

\textbf{Distraction Level.} This metric reflects the extent to which an icon's position can divert the attention of the driver, with higher scores indicating greater distraction. Significant differences are observed in simulated driving, where P2 (middle-center) has the highest distraction score of 5.68, distinguishing it from P1 (top-center, P=0.000), P3 (bottom-center, P=0.004), P4 (top-right, P=0.000), and P5 (middle-right, P=0.000). A difference between P1 and P3 (P=0.019) is also noted. In real-world navigation, P2, with the highest score of 5.80, differs significantly from P1 (top-center, P=0.000), P3 (bottom-center, P=0.017), P4 (top-right, P=0.002), and P5 (center-right, P=0.017). The \textbf{top-right (P4)} receives the lowest distraction scores in both simulated and real-world driving, with 2.43 and 2.30, respectively, indicating it as the least distracting location, while the second best is both \textbf{top-center (P1)}, with scores of 2.71 in simulated driving and 2.75 in real-world navigation.

\textbf{Workload Level.} This metric measures both mental and physical demands associated with the navigation icon position, where higher scores indicate more effort required from the participants. In the simulated driving experiment, significant differences are observed, with P4 (top-right) demanding the most attention, reflected by the highest score of 3.83 when compared to P1 (top-center, P=0.007), P2 (middle-center, P=0.000), P3 (bottom-center, P=0.000), and P5 (middle-right, P=0.013). Additionally, P2 with the lowest score of 1.53, is significantly less demanding than P5 (P=0.017). In the real-world navigation, P3 (bottom-center) shows a significantly less workload score of 1.65 when compared to P1 (top-center, P=0.010), P4 (top-right, P=0.005), and P5 (middle-right, P=0.049), with P2 (middle-center) also differing significantly from P1 (P=0.001), and P4 (P=0.001). \textbf{middle-center (P2)} consistently shows the lowest workload scores, 1.53 in simulated and 1.45 in real-world driving, suggesting it is the least demanding position. Additionally, \textbf{bottom-center (P3)} demonstrates relatively lower workload levels in both settings.

\subsection{User Preference Interview}

Following the simulated driving and real-world navigation experiment, participants were asked to identify the optimal and least favorable positions, in addition to providing their opinions. This exercise aimed to gather subjective feedback on the preferences based on their past experiences of driving, which could offer valuable insights into the design and evaluation of the AR interfaces or navigation solutions. The result is summarized in Table \ref{table:preferences}.

\begin{table}[htbp]
\centering
\caption{Summary of Participant Preferences for Icon Locations}
\label{table:preferences}
\begin{tabular}{p{0.20\linewidth} p{0.35\linewidth} p{0.35\linewidth}}
\toprule
\textbf{Position} & \textbf{Simulated Driving Preferences} & \textbf{Real-world Navigation Preferences} \\
\midrule
Top-Center (P1) & Favored by 7 participants for its prominence and alignment with habitual driving vision, allowing \textbf{easy glances without obstructing the view}. & Chosen by only 1 participant, criticized for \textbf{conflicting with traffic signs and lights} due to the unnatural upward eye movement required. \\
Middle-Center (P2) & Least acceptable to 11 participants due to its \textbf{obstruction of road visibility and significant safety risks}.  & Deemed least suitable by 7 participants, often \textbf{overlapping with the vehicle ahead and causing view obstruction}. \\
Bottom-Center (P3) & Preferred by 1 participant for its central location; criticized for \textbf{instability during head movements and middle position obstructing the road view}. & Preferred by 6 and least by 1; considered \textbf{minimally intrusive} to windshield visibility but suggested to need \textbf{stability when moving horizontally}. \\
Top-Right (P4) & Preferred by 3 participants for causing \textbf{less distraction than central placements}; 1 found it \textbf{too distant and effortful} to check. & Found most suitable by 3 participants for \textbf{minimal distraction}; however, 2 noted it was \textbf{too far from the usual field of vision}, requiring more effort to view. \\
Center-Right (P5) & Chosen by 1 participant for its \textbf{stability and non-obstructive placement} compared to top and bottom positions. & Received mixed reactions; some noted it \textbf{obstructed views of the right-hand road scene}, while others found it \textbf{discreet and particularly noticeable at night} due to contrasting lighting.\\
\bottomrule
\end{tabular}
\end{table}

\section{Discussion}

\subsection{Preference for Information Placement}

\begin{figure}[htbp]
  \centering
  \includegraphics[width=0.65\linewidth]{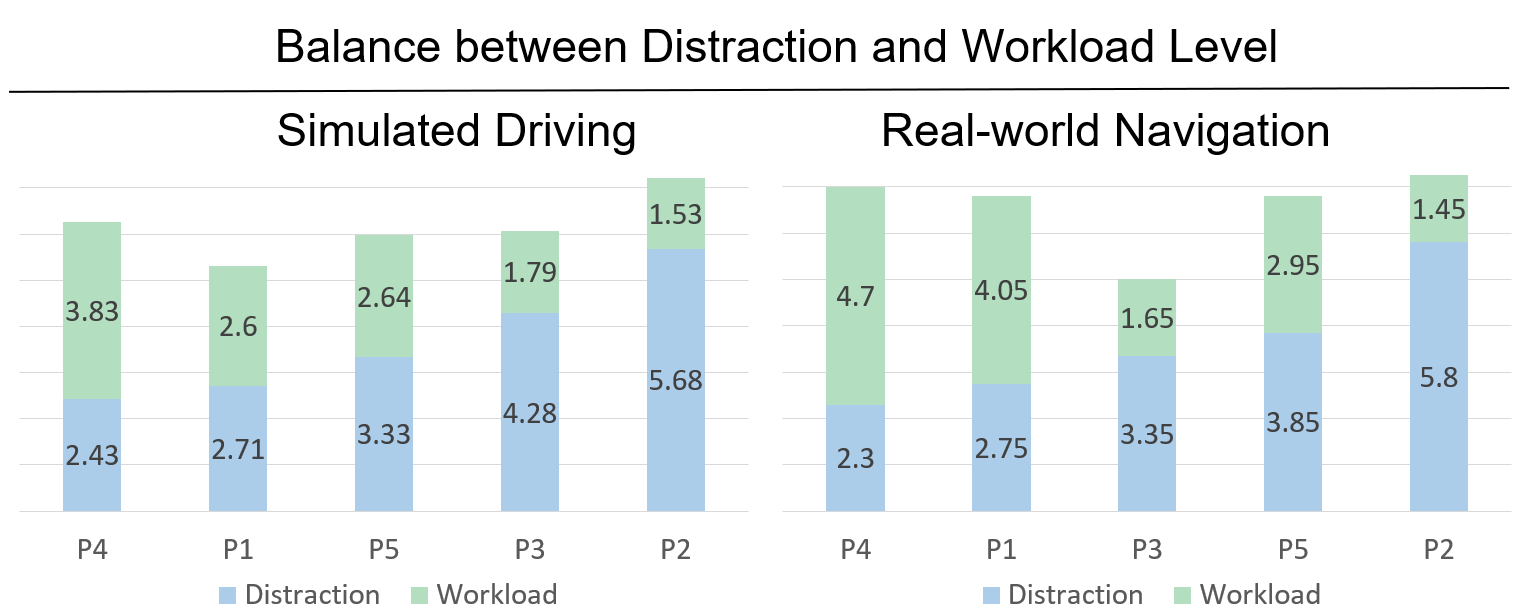}
  \caption{Cumulative bar chart of Distraction and Workload Level for simulated driving and real-world navigation scenarios, sorted by Distraction Level respectively.}
  \Description{The bar chart compares the balance between Distraction and Workload levels across participants in simulated driving and real-world navigation scenarios. For both scenarios, bars are divided into two segments representing Distraction (in green) and Workload (in blue), with values provided for each. Icon positions are labeled from P1 to P5.}
  \label{fig:balance}
\end{figure}

According to the experiment results, the central position not only brings the lowest workload to participants but also brings the highest distraction level. While top-right is just the opposite, it brings the lowest distraction, but the highest workload. Based on this intuitive finding, we add up the average score of distraction level and workload level together for two scenarios, which is shown in Figure \ref{fig:balance}. The position name of the icon is sorted in ascending order based on the distraction level. The figure reveals that in both scenarios, as the level of distraction increases, there tends to be a decrease in the workload level, suggesting an interaction effect between the two factors. Further analysis indicates that in simulated driving scenarios, \textbf{top-center (P1)} has the lowest cumulative score (5.31) of the two factors, suggesting its superiority for simulated environments. Conversely, in real-world navigation scenarios, \textbf{bottom-center (P3)} exhibits the lowest cumulative score (5.00), indicating its optimal suitability under real-world conditions. \textbf{Middle-center (P2)} records the highest cumulative scores in both scenarios (7.21 and 7.25), corresponding with its minimal overall appropriateness. This finding aligns well with the information placement analysis and user preference interview, where most participants prefer the top-center in simulated environments and the bottom-center in real-world scenarios, with significant differences from other positions. This study underscores a fundamental aspect of AR interface design: \textit{the necessity to find a balance between distraction and workload}. 

The discrepancy in preferred locations between simulated and real-world settings can be attributed to the complexities and unpredictability inherent in real-world driving as well as visual-related ergonomic factors. Different from simulation, there exist many traffic signs hanging high in the air which might cost more attention for drivers to distinguish with the navigation icon. This kind of traffic sign was merely taken into consideration in past research. Also, outdoor lighting conditions were more complex and difficult to replicate through simulation devices. The visual qualities such as contrast might decrease a lot in outdoor usage when participants saw-through the display with the sky as the background. Unlike the controlled simulated environment, where drivers could afford to glance upward more safely, the real-world environment demands relatively higher and more continuous attention to the central vision field. Also, lower placements that are closer to the road level might better align with the natural line of eyesight and driving practices, as evidenced by the preference for the bottom-center position. This finding also aligns with Rzayev et al. \cite{rzayev2018reading}, Imamov et al. \cite{imamov2020display}, and Chua et al. \cite{chua2016positioning}, they find that the bottom-center is a preferred position in their experimental setups. These findings also corroborate prior research indicating that drivers (and AR designers) should avoid tasks that require drivers to look away from the forward roadway \cite{olson2009driver}, indicating the importance of real-world tests. Our study contributes to this body of knowledge by quantifying the impact of specific icon placements with a portable AR device in a realistic context, suggesting that the bottom-center location would be a relatively better choice for future AR navigation systems.

\subsection{Design Implications}

This research highlights four pivotal aspects for designing AR interfaces in automotive applications: visual ergonomics, lighting conditions, traffic elements, and body movements. These components are essential for creating AR navigation systems, ultimately improving the safety and efficiency of the driving experience.

\textbf{Visual Ergonomics.} The findings present the critical role of ergonomic principles in the design of AR interfaces, with an emphasis on user comfort and the natural patterns of interaction. It is crucial to strategically position visual information to ensure it does not impede the view of the driver while remaining accessible at a glance. According to human factors research, the optimal angle for easy eye rotation is approximately 30 degrees downward, whereas the natural resting line of sight is around 15 degrees downward \cite{kress2019digital,dreyfuss1993measure}. These guidelines corroborate our results in real-world conditions, underscoring a design approach that delivers essential information without inducing physical overload.

\textbf{Lighting Conditions.} The impact of environmental lighting on the perceived quality of information is substantial and can significantly affect workload during interface interaction. Gabbard et al. \cite{6549410} demonstrate that in AR systems, specifically through optical see-through head-mounted displays (oHMDs), the colors of AR elements become faded under high luminance backgrounds. This effect is particularly noticeable given that the brightness of the sky is typically higher than that of the ground, resulting in lower contrast for human eyes. Therefore, it is essential to test factual visual quality under a variety of environmental and lighting conditions to ensure its validity in real-world applications.

\textbf{Traffic Elements.} Real-driving scenarios encompass a diverse array of traffic components, including not only signs and moving vehicles but also pedestrians and static obstacles. It is imperative to avoid overlaying navigation information on these crucial elements. Maintaining precise 2D and 3D spatial relationships between AR elements and these traffic components is essential to prevent visual occlusion and confusion.

\textbf{Body Movements.} AR interface must be highly adaptable and intuitive, particularly in accommodating a driver's body movements during driving. As drivers adjust their posture and head orientation in response to road conditions, it is crucial for AR information to remain stable and clear by design or technical methods. This ensures that essential navigation and safety information is consistently accessible and properly aligned with the driver’s line of sight. Dynamic adjustment of AR elements to body movements is key to maintaining driver trust and preventing cognitive confusion.

\section{Limitations and Future Work}

This study, while providing valuable insights, has limitations that warrant further investigation. The sample size and demographic diversity of participants, need expansion to better represent the diverse driving population. Additionally, the exploration of AR navigation in different environmental conditions remains essential to refine context-aware interface designs. Longitudinal studies are also imperative to understand the sustained effects of AR navigation on driving behavior and safety.

Although drivers prefer non-central placement for AR navigation, in special contexts like autonomous driving where immediate take-over requests (TORs) \cite{10.1145/3003715.3005409} may be necessitated, centrally positioned icons could prove more effective in averting accidents. While the current research emphasizes the visual components of AR interfaces, incorporating multi-modal feedback could offer a more comprehensive understanding of user interaction in driving contexts. Exploring these additional feedback mechanisms may help in reducing the visual load and enhancing the user experience with AR navigation systems. The development of adaptive AR systems, utilizing machine learning to dynamically align with user needs and driving contexts, stands as an essential next step in advancing AR navigation technologies.

\section{Conclusion}

In conclusion, this study investigates the impact of information placement on the user interface of AR glasses for driving applications. Utilizing an AR device that exceeds traditional models in terms of lightness and brightness, the research offers practical insights for future automotive innovations. Through experiments conducted in simulated and real-world driving settings, the study determines optimal placements: top-center in simulated environments and bottom-center in real-world conditions. The similarity and divergence between the two scenarios not only reveal a consistent pattern of balance between distraction and workload but also highlight the crucial role of real-world experimentation in refining AR interface design. The integration of qualitative insights and quantitative data underscores four design implications, emphasizing AR navigation systems that are intuitive, non-intrusive, adaptable, and tailored to a variety of driving scenarios and user preferences. These findings contribute to the advancement of AR navigation technologies applied in automotive area, with a focus on ergonomic and practical enhancements that improve the driving experience.

\begin{acks}
The authors wish to thank the OPPO Research Institute for supporting this research.
\end{acks}

\bibliographystyle{ACM-Reference-Format}
\bibliography{sample-base}


\begin{thebibliography}{49}


\ifx \showCODEN    \undefined \def \showCODEN     #1{\unskip}     \fi
\ifx \showDOI      \undefined \def \showDOI       #1{#1}\fi
\ifx \showISBNx    \undefined \def \showISBNx     #1{\unskip}     \fi
\ifx \showISBNxiii \undefined \def \showISBNxiii  #1{\unskip}     \fi
\ifx \showISSN     \undefined \def \showISSN      #1{\unskip}     \fi
\ifx \showLCCN     \undefined \def \showLCCN      #1{\unskip}     \fi
\ifx \shownote     \undefined \def \shownote      #1{#1}          \fi
\ifx \showarticletitle \undefined \def \showarticletitle #1{#1}   \fi
\ifx \showURL      \undefined \def \showURL       {\relax}        \fi
\providecommand\bibfield[2]{#2}
\providecommand\bibinfo[2]{#2}
\providecommand\natexlab[1]{#1}
\providecommand\showeprint[2][]{arXiv:#2}

\bibitem[Alves et~al\mbox{.}(2013)]%
        {alves2013forward}
\bibfield{author}{\bibinfo{person}{Patr{\'\i}cia~RJA Alves}, \bibinfo{person}{Joel Gon{\c{c}}alves}, \bibinfo{person}{Rosaldo~JF Rossetti}, \bibinfo{person}{Eug{\'e}nio~C Oliveira}, {and} \bibinfo{person}{Cristina Olaverri-Monreal}.} \bibinfo{year}{2013}\natexlab{}.
\newblock \showarticletitle{Forward collision warning systems using heads-up displays: Testing usability of two new metaphors}. In \bibinfo{booktitle}{\emph{2013 IEEE Intelligent Vehicles Symposium Workshops (IV Workshops)}}. IEEE, \bibinfo{pages}{1--6}.
\newblock


\bibitem[Anderson et~al\mbox{.}(2019)]%
        {9020354}
\bibfield{author}{\bibinfo{person}{Ryan Anderson}, \bibinfo{person}{Juan Toledo}, {and} \bibinfo{person}{Hala ElAarag}.} \bibinfo{year}{2019}\natexlab{}.
\newblock \showarticletitle{Feasibility Study on the Utilization of Microsoft HoloLens to Increase Driving Conditions Awareness}. In \bibinfo{booktitle}{\emph{2019 SoutheastCon}}. \bibinfo{pages}{1--8}.
\newblock
\urldef\tempurl%
\url{https://doi.org/10.1109/SoutheastCon42311.2019.9020354}
\showDOI{\tempurl}


\bibitem[Astrologo et~al\mbox{.}(2024)]%
        {astrologo2024determining}
\bibfield{author}{\bibinfo{person}{Amanda~N Astrologo}, \bibinfo{person}{Sarah Nano}, \bibinfo{person}{Elizabeth~M Klemm}, \bibinfo{person}{Sandra~J Shefelbine}, {and} \bibinfo{person}{Jack~T Dennerlein}.} \bibinfo{year}{2024}\natexlab{}.
\newblock \showarticletitle{Determining the effects of AR/VR HMD design parameters (mass and inertia) on cervical spine joint torques}.
\newblock \bibinfo{journal}{\emph{Applied Ergonomics}}  \bibinfo{volume}{116} (\bibinfo{year}{2024}), \bibinfo{pages}{104183}.
\newblock


\bibitem[Bauerfeind et~al\mbox{.}(2021)]%
        {bauerfeind2021navigating}
\bibfield{author}{\bibinfo{person}{Kassandra Bauerfeind}, \bibinfo{person}{Julia Dr{\"u}ke}, \bibinfo{person}{Jens Schneider}, \bibinfo{person}{Adrian Haar}, \bibinfo{person}{Lennart Bendewald}, {and} \bibinfo{person}{Martin Baumann}.} \bibinfo{year}{2021}\natexlab{}.
\newblock \showarticletitle{Navigating with Augmented Reality--How does it affect drivers’ mental load?}
\newblock \bibinfo{journal}{\emph{Applied ergonomics}}  \bibinfo{volume}{94} (\bibinfo{year}{2021}), \bibinfo{pages}{103398}.
\newblock


\bibitem[Bell et~al\mbox{.}(2002)]%
        {bell2002information}
\bibfield{author}{\bibinfo{person}{Blaine Bell}, \bibinfo{person}{Steven Feiner}, {and} \bibinfo{person}{Tobias Hollerer}.} \bibinfo{year}{2002}\natexlab{}.
\newblock \showarticletitle{Information at a glance [augmented reality user interfaces]}.
\newblock \bibinfo{journal}{\emph{IEEE Computer Graphics and Applications}} \bibinfo{volume}{22}, \bibinfo{number}{4} (\bibinfo{year}{2002}), \bibinfo{pages}{6--9}.
\newblock


\bibitem[Bolder et~al\mbox{.}(2018)]%
        {bolder2018comparison}
\bibfield{author}{\bibinfo{person}{Anna Bolder}, \bibinfo{person}{Stefan~M Gr{\"u}nvogel}, {and} \bibinfo{person}{Emanuel Angelescu}.} \bibinfo{year}{2018}\natexlab{}.
\newblock \showarticletitle{Comparison of the usability of a car infotainment system in a mixed reality environment and in a real car}. In \bibinfo{booktitle}{\emph{Proceedings of the 24th ACM Symposium on Virtual Reality Software and Technology}}. \bibinfo{pages}{1--10}.
\newblock


\bibitem[Bolton et~al\mbox{.}(2015)]%
        {10.1145/2799250.2799253}
\bibfield{author}{\bibinfo{person}{Adam Bolton}, \bibinfo{person}{Gary Burnett}, {and} \bibinfo{person}{David~R Large}.} \bibinfo{year}{2015}\natexlab{}.
\newblock \showarticletitle{An investigation of augmented reality presentations of landmark-based navigation using a head-up display}. In \bibinfo{booktitle}{\emph{Proceedings of the 7th International Conference on Automotive User Interfaces and Interactive Vehicular Applications}} (Nottingham, United Kingdom) \emph{(\bibinfo{series}{AutomotiveUI '15})}. \bibinfo{publisher}{Association for Computing Machinery}, \bibinfo{address}{New York, NY, USA}, \bibinfo{pages}{56–63}.
\newblock
\showISBNx{9781450337366}
\urldef\tempurl%
\url{https://doi.org/10.1145/2799250.2799253}
\showDOI{\tempurl}


\bibitem[Borojeni et~al\mbox{.}(2016)]%
        {10.1145/3003715.3005409}
\bibfield{author}{\bibinfo{person}{Shadan~Sadeghian Borojeni}, \bibinfo{person}{Lewis Chuang}, \bibinfo{person}{Wilko Heuten}, {and} \bibinfo{person}{Susanne Boll}.} \bibinfo{year}{2016}\natexlab{}.
\newblock \showarticletitle{Assisting Drivers with Ambient Take-Over Requests in Highly Automated Driving}. In \bibinfo{booktitle}{\emph{Proceedings of the 8th International Conference on Automotive User Interfaces and Interactive Vehicular Applications}} (Ann Arbor, MI, USA) \emph{(\bibinfo{series}{Automotive'UI 16})}. \bibinfo{publisher}{Association for Computing Machinery}, \bibinfo{address}{New York, NY, USA}, \bibinfo{pages}{237–244}.
\newblock
\showISBNx{9781450345330}
\urldef\tempurl%
\url{https://doi.org/10.1145/3003715.3005409}
\showDOI{\tempurl}


\bibitem[Bram-Larbi et~al\mbox{.}(2020)]%
        {bram2020collision}
\bibfield{author}{\bibinfo{person}{Kweku~F Bram-Larbi}, \bibinfo{person}{Vassilis Charissis}, \bibinfo{person}{Soheeb Khan}, \bibinfo{person}{Ramesh Lagoo}, \bibinfo{person}{David~K Harrison}, {and} \bibinfo{person}{Dimitris Drikakis}.} \bibinfo{year}{2020}\natexlab{}.
\newblock \showarticletitle{Collision avoidance head-up display: design considerations for emergency services’ vehicles}. In \bibinfo{booktitle}{\emph{2020 IEEE International Conference on Consumer Electronics (ICCE)}}. IEEE, \bibinfo{pages}{1--7}.
\newblock


\bibitem[Calvi et~al\mbox{.}(2020)]%
        {calvi2020effectiveness}
\bibfield{author}{\bibinfo{person}{Alessandro Calvi}, \bibinfo{person}{Fabrizio D’Amico}, \bibinfo{person}{Chiara Ferrante}, {and} \bibinfo{person}{Luca~Bianchini Ciampoli}.} \bibinfo{year}{2020}\natexlab{}.
\newblock \showarticletitle{Effectiveness of augmented reality warnings on driving behaviour whilst approaching pedestrian crossings: A driving simulator study}.
\newblock \bibinfo{journal}{\emph{Accident Analysis \& Prevention}}  \bibinfo{volume}{147} (\bibinfo{year}{2020}), \bibinfo{pages}{105760}.
\newblock


\bibitem[Charissis(2014)]%
        {charissis2014enhancing}
\bibfield{author}{\bibinfo{person}{Vassilis Charissis}.} \bibinfo{year}{2014}\natexlab{}.
\newblock \showarticletitle{Enhancing human responses through augmented reality Head-Up Display in vehicular environment}. In \bibinfo{booktitle}{\emph{2014 11th International Conference \& Expo on Emerging Technologies for a Smarter World (CEWIT)}}. IEEE, \bibinfo{pages}{1--6}.
\newblock


\bibitem[Chen et~al\mbox{.}(2017)]%
        {chen2017eliminating}
\bibfield{author}{\bibinfo{person}{Bo-Hao Chen}, \bibinfo{person}{Shih-Chia Huang}, {and} \bibinfo{person}{Wei-Ho Tsai}.} \bibinfo{year}{2017}\natexlab{}.
\newblock \showarticletitle{Eliminating driving distractions: Human-computer interaction with built-in applications}.
\newblock \bibinfo{journal}{\emph{IEEE Vehicular Technology Magazine}} \bibinfo{volume}{12}, \bibinfo{number}{1} (\bibinfo{year}{2017}), \bibinfo{pages}{20--29}.
\newblock


\bibitem[Chen et~al\mbox{.}(2021)]%
        {chen2021human}
\bibfield{author}{\bibinfo{person}{Yumiao Chen}, \bibinfo{person}{Xin Wang}, {and} \bibinfo{person}{Huijia Xu}.} \bibinfo{year}{2021}\natexlab{}.
\newblock \showarticletitle{Human factors/ergonomics evaluation for virtual reality headsets: a review}.
\newblock \bibinfo{journal}{\emph{CCF Transactions on Pervasive Computing and Interaction}} \bibinfo{volume}{3}, \bibinfo{number}{2} (\bibinfo{year}{2021}), \bibinfo{pages}{99--111}.
\newblock


\bibitem[Chua et~al\mbox{.}(2016)]%
        {chua2016positioning}
\bibfield{author}{\bibinfo{person}{Soon~Hau Chua}, \bibinfo{person}{Simon~T Perrault}, \bibinfo{person}{Denys~JC Matthies}, {and} \bibinfo{person}{Shengdong Zhao}.} \bibinfo{year}{2016}\natexlab{}.
\newblock \showarticletitle{Positioning glass: Investigating display positions of monocular optical see-through head-mounted display}. In \bibinfo{booktitle}{\emph{Proceedings of the Fourth International Symposium of Chinese CHI}}. \bibinfo{pages}{1--6}.
\newblock


\bibitem[Detjen et~al\mbox{.}(2022)]%
        {10.1145/3543174.3546089}
\bibfield{author}{\bibinfo{person}{Henrik Detjen}, \bibinfo{person}{Sarah Faltaous}, \bibinfo{person}{Jonas Keppel}, \bibinfo{person}{Marvin Prochazka}, \bibinfo{person}{Uwe Gruenefeld}, \bibinfo{person}{Shadan Sadeghian}, {and} \bibinfo{person}{Stefan Schneegass}.} \bibinfo{year}{2022}\natexlab{}.
\newblock \showarticletitle{Investigating the Influence of Gaze- and Context-Adaptive Head-up Displays on Take-Over Requests}. In \bibinfo{booktitle}{\emph{Proceedings of the 14th International Conference on Automotive User Interfaces and Interactive Vehicular Applications}} (Seoul, Republic of Korea) \emph{(\bibinfo{series}{AutomotiveUI '22})}. \bibinfo{publisher}{Association for Computing Machinery}, \bibinfo{address}{New York, NY, USA}, \bibinfo{pages}{108–118}.
\newblock
\showISBNx{9781450394154}
\urldef\tempurl%
\url{https://doi.org/10.1145/3543174.3546089}
\showDOI{\tempurl}


\bibitem[Dreyfuss et~al\mbox{.}(1993)]%
        {dreyfuss1993measure}
\bibfield{author}{\bibinfo{person}{Henry Dreyfuss}, \bibinfo{person}{Henry~Dreyfuss Associates}, {and} \bibinfo{person}{Alvin~R Tilley}.} \bibinfo{year}{1993}\natexlab{}.
\newblock \bibinfo{booktitle}{\emph{The measure of man and woman: human factors in design}}.
\newblock \bibinfo{publisher}{Whitney Library of Design}.
\newblock


\bibitem[Du et~al\mbox{.}(2023)]%
        {du2023comfort}
\bibfield{author}{\bibinfo{person}{Yujia Du}, \bibinfo{person}{Kexiang Liu}, \bibinfo{person}{Yuxin Ju}, {and} \bibinfo{person}{Haining Wang}.} \bibinfo{year}{2023}\natexlab{}.
\newblock \showarticletitle{A comfort analysis of AR glasses on physical load during long-term wearing}.
\newblock \bibinfo{journal}{\emph{Ergonomics}} \bibinfo{volume}{66}, \bibinfo{number}{9} (\bibinfo{year}{2023}), \bibinfo{pages}{1325--1339}.
\newblock


\bibitem[Franchak et~al\mbox{.}(2021)]%
        {franchak2021adapting}
\bibfield{author}{\bibinfo{person}{John~M Franchak}, \bibinfo{person}{Brianna McGee}, {and} \bibinfo{person}{Gabrielle Blanch}.} \bibinfo{year}{2021}\natexlab{}.
\newblock \showarticletitle{Adapting the coordination of eyes and head to differences in task and environment during fully-mobile visual exploration}.
\newblock \bibinfo{journal}{\emph{PLoS one}} \bibinfo{volume}{16}, \bibinfo{number}{8} (\bibinfo{year}{2021}), \bibinfo{pages}{e0256463}.
\newblock


\bibitem[Fr{\'e}mont et~al\mbox{.}(2020)]%
        {fremont2020adaptive}
\bibfield{author}{\bibinfo{person}{Vincent Fr{\'e}mont}, \bibinfo{person}{Minh-Tien Phan}, {and} \bibinfo{person}{Indira Thouvenin}.} \bibinfo{year}{2020}\natexlab{}.
\newblock \showarticletitle{Adaptive visual assistance system for enhancing the driver awareness of pedestrians}.
\newblock \bibinfo{journal}{\emph{International Journal of Human--Computer Interaction}} \bibinfo{volume}{36}, \bibinfo{number}{9} (\bibinfo{year}{2020}), \bibinfo{pages}{856--869}.
\newblock


\bibitem[Gabbard et~al\mbox{.}(2014)]%
        {gabbard2014behind}
\bibfield{author}{\bibinfo{person}{Joseph~L Gabbard}, \bibinfo{person}{Gregory~M Fitch}, {and} \bibinfo{person}{Hyungil Kim}.} \bibinfo{year}{2014}\natexlab{}.
\newblock \showarticletitle{Behind the glass: Driver challenges and opportunities for AR automotive applications}.
\newblock \bibinfo{journal}{\emph{Proc. IEEE}} \bibinfo{volume}{102}, \bibinfo{number}{2} (\bibinfo{year}{2014}), \bibinfo{pages}{124--136}.
\newblock


\bibitem[Gabbard et~al\mbox{.}(2019)]%
        {gabbard2019ar}
\bibfield{author}{\bibinfo{person}{Joseph~L Gabbard}, \bibinfo{person}{Missie Smith}, \bibinfo{person}{Kyle Tanous}, \bibinfo{person}{Hyungil Kim}, {and} \bibinfo{person}{Bryan Jonas}.} \bibinfo{year}{2019}\natexlab{}.
\newblock \showarticletitle{AR drivesim: An immersive driving simulator for augmented reality head-up display research}.
\newblock \bibinfo{journal}{\emph{Frontiers in Robotics and AI}}  \bibinfo{volume}{6} (\bibinfo{year}{2019}), \bibinfo{pages}{98}.
\newblock


\bibitem[Gabbard et~al\mbox{.}(2013)]%
        {6549410}
\bibfield{author}{\bibinfo{person}{Joseph~L. Gabbard}, \bibinfo{person}{J.~Edward Swan}, {and} \bibinfo{person}{Adam Zarger}.} \bibinfo{year}{2013}\natexlab{}.
\newblock \showarticletitle{Color blending in outdoor optical see-through AR: The effect of real-world backgrounds on user interface color}. In \bibinfo{booktitle}{\emph{2013 IEEE Virtual Reality (VR)}}. \bibinfo{pages}{157--158}.
\newblock
\urldef\tempurl%
\url{https://doi.org/10.1109/VR.2013.6549410}
\showDOI{\tempurl}


\bibitem[Haeuslschmid et~al\mbox{.}(2016)]%
        {haeuslschmid2016design}
\bibfield{author}{\bibinfo{person}{Renate Haeuslschmid}, \bibinfo{person}{Bastian Pfleging}, {and} \bibinfo{person}{Florian Alt}.} \bibinfo{year}{2016}\natexlab{}.
\newblock \showarticletitle{A design space to support the development of windshield applications for the car}. In \bibinfo{booktitle}{\emph{Proceedings of the 2016 CHI Conference on Human Factors in Computing Systems}}. \bibinfo{pages}{5076--5091}.
\newblock


\bibitem[Horrey et~al\mbox{.}(2003)]%
        {horrey2003effects}
\bibfield{author}{\bibinfo{person}{William~J Horrey}, \bibinfo{person}{Christopher~D Wickens}, {and} \bibinfo{person}{Amy~L Alexander}.} \bibinfo{year}{2003}\natexlab{}.
\newblock \showarticletitle{The effects of head-up display clutter and in-vehicle display separation on concurrent driving performance}. In \bibinfo{booktitle}{\emph{Proceedings of the Human Factors and Ergonomics Society Annual Meeting}}, Vol.~\bibinfo{volume}{47}. SAGE Publications Sage CA: Los Angeles, CA, \bibinfo{pages}{1880--1884}.
\newblock


\bibitem[Imamov et~al\mbox{.}(2020)]%
        {imamov2020display}
\bibfield{author}{\bibinfo{person}{Samat Imamov}, \bibinfo{person}{Daniel Monzel}, {and} \bibinfo{person}{Wallace~S Lages}.} \bibinfo{year}{2020}\natexlab{}.
\newblock \showarticletitle{Where to display? how interface position affects comfort and task switching time on glanceable interfaces}. In \bibinfo{booktitle}{\emph{2020 IEEE Conference on Virtual Reality and 3D User Interfaces (VR)}}. IEEE, \bibinfo{pages}{851--858}.
\newblock


\bibitem[Kemeny(2023)]%
        {kemeny2023virtual}
\bibfield{author}{\bibinfo{person}{Andras Kemeny}.} \bibinfo{year}{2023}\natexlab{}.
\newblock \showarticletitle{Virtual and Augmented Reality}.
\newblock In \bibinfo{booktitle}{\emph{Autonomous Vehicles and Virtual Reality: The New Automobile Industrial Revolution}}. \bibinfo{publisher}{Springer}, \bibinfo{pages}{33--49}.
\newblock


\bibitem[Kim and Gabbard(2022)]%
        {kim2022assessing}
\bibfield{author}{\bibinfo{person}{Hyungil Kim} {and} \bibinfo{person}{Joseph~L Gabbard}.} \bibinfo{year}{2022}\natexlab{}.
\newblock \showarticletitle{Assessing distraction potential of augmented reality head-up displays for vehicle drivers}.
\newblock \bibinfo{journal}{\emph{Human factors}} \bibinfo{volume}{64}, \bibinfo{number}{5} (\bibinfo{year}{2022}), \bibinfo{pages}{852--865}.
\newblock


\bibitem[Kim et~al\mbox{.}(2013)]%
        {kim2013exploring}
\bibfield{author}{\bibinfo{person}{Hyungil Kim}, \bibinfo{person}{Xuefang Wu}, \bibinfo{person}{Joseph~L Gabbard}, {and} \bibinfo{person}{Nicholas~F Polys}.} \bibinfo{year}{2013}\natexlab{}.
\newblock \showarticletitle{Exploring head-up augmented reality interfaces for crash warning systems}. In \bibinfo{booktitle}{\emph{Proceedings of the 5th International Conference on Automotive User Interfaces and Interactive Vehicular Applications}}. \bibinfo{pages}{224--227}.
\newblock


\bibitem[Kress(2019)]%
        {kress2019digital}
\bibfield{author}{\bibinfo{person}{Bernard~C Kress}.} \bibinfo{year}{2019}\natexlab{}.
\newblock \showarticletitle{Digital optical elements and technologies (EDO19): applications to AR/VR/MR}. In \bibinfo{booktitle}{\emph{Digital Optical Technologies 2019}}, Vol.~\bibinfo{volume}{11062}. SPIE, \bibinfo{pages}{343--355}.
\newblock


\bibitem[Kun et~al\mbox{.}(2017)]%
        {kun2017calling}
\bibfield{author}{\bibinfo{person}{Andrew~L Kun}, \bibinfo{person}{Steven~W van~der Meulen}, {and} \bibinfo{person}{Christian~P Janssen}.} \bibinfo{year}{2017}\natexlab{}.
\newblock \showarticletitle{Calling while driving: An initial experiment with HoloLens}. In \bibinfo{booktitle}{\emph{Driving Assessment Conference}}, Vol.~\bibinfo{volume}{9}. University of Iowa.
\newblock


\bibitem[Lee and Woo(2022)]%
        {lee2022exploring}
\bibfield{author}{\bibinfo{person}{Hyunjin Lee} {and} \bibinfo{person}{Woontack Woo}.} \bibinfo{year}{2022}\natexlab{}.
\newblock \showarticletitle{Exploring Augmented Reality Notification Placement while Communicating with Virtual Avatar}. In \bibinfo{booktitle}{\emph{2022 IEEE International Symposium on Mixed and Augmented Reality Adjunct (ISMAR-Adjunct)}}. IEEE, \bibinfo{pages}{686--689}.
\newblock


\bibitem[Li et~al\mbox{.}(2020)]%
        {li2020improving}
\bibfield{author}{\bibinfo{person}{Quanyi Li}, \bibinfo{person}{Zhenghao Peng}, \bibinfo{person}{Qihang Zhang}, \bibinfo{person}{Chunxiao Liu}, {and} \bibinfo{person}{Bolei Zhou}.} \bibinfo{year}{2020}\natexlab{}.
\newblock \showarticletitle{Improving the generalization of end-to-end driving through procedural generation}.
\newblock \bibinfo{journal}{\emph{arXiv preprint arXiv:2012.13681}} (\bibinfo{year}{2020}).
\newblock


\bibitem[Ma et~al\mbox{.}(2021)]%
        {ma2021impact}
\bibfield{author}{\bibinfo{person}{Xiangdong Ma}, \bibinfo{person}{Zhicong Hong}, \bibinfo{person}{Junhong Huang}, \bibinfo{person}{Jingjing He}, \bibinfo{person}{Jianpeng Song}, {and} \bibinfo{person}{Mengting Jia}.} \bibinfo{year}{2021}\natexlab{}.
\newblock \showarticletitle{The impact of AR-HUD intelligent driving on the allocation of cognitive resources under the breakthrough of 5G technology}. In \bibinfo{booktitle}{\emph{Journal of Physics: Conference Series}}, Vol.~\bibinfo{volume}{1982}. IOP Publishing, \bibinfo{pages}{012024}.
\newblock


\bibitem[Manger et~al\mbox{.}(2023)]%
        {10.1145/3581961.3609874}
\bibfield{author}{\bibinfo{person}{Carina Manger}, \bibinfo{person}{Jakob Peintner}, \bibinfo{person}{Marion Hoffmann}, \bibinfo{person}{Mirella Probst}, \bibinfo{person}{Raphael Wennmacher}, {and} \bibinfo{person}{Andreas Riener}.} \bibinfo{year}{2023}\natexlab{}.
\newblock \showarticletitle{Providing Explainability in Safety-Critical Automated Driving Situations through Augmented Reality Windshield HMIs}. In \bibinfo{booktitle}{\emph{Adjunct Proceedings of the 15th International Conference on Automotive User Interfaces and Interactive Vehicular Applications}} (Ingolstadt, Germany) \emph{(\bibinfo{series}{AutomotiveUI '23 Adjunct})}. \bibinfo{publisher}{Association for Computing Machinery}, \bibinfo{address}{New York, NY, USA}, \bibinfo{pages}{174–179}.
\newblock
\showISBNx{9798400701122}
\urldef\tempurl%
\url{https://doi.org/10.1145/3581961.3609874}
\showDOI{\tempurl}


\bibitem[Medenica et~al\mbox{.}(2011)]%
        {medenica2011augmented}
\bibfield{author}{\bibinfo{person}{Zeljko Medenica}, \bibinfo{person}{Andrew~L Kun}, \bibinfo{person}{Tim Paek}, {and} \bibinfo{person}{Oskar Palinko}.} \bibinfo{year}{2011}\natexlab{}.
\newblock \showarticletitle{Augmented reality vs. street views: a driving simulator study comparing two emerging navigation aids}. In \bibinfo{booktitle}{\emph{Proceedings of the 13th International Conference on Human Computer Interaction with Mobile Devices and Services}}. \bibinfo{pages}{265--274}.
\newblock


\bibitem[Mekni and Lemieux(2014)]%
        {mekni2014augmented}
\bibfield{author}{\bibinfo{person}{Mehdi Mekni} {and} \bibinfo{person}{Andre Lemieux}.} \bibinfo{year}{2014}\natexlab{}.
\newblock \showarticletitle{Augmented reality: Applications, challenges and future trends}.
\newblock \bibinfo{journal}{\emph{Applied computational science}}  \bibinfo{volume}{20} (\bibinfo{year}{2014}), \bibinfo{pages}{205--214}.
\newblock


\bibitem[Nwakacha et~al\mbox{.}(2013)]%
        {nwakacha2013evaluating}
\bibfield{author}{\bibinfo{person}{Valentine Nwakacha}, \bibinfo{person}{Andy Crabtree}, {and} \bibinfo{person}{Gary Burnett}.} \bibinfo{year}{2013}\natexlab{}.
\newblock \showarticletitle{Evaluating distraction and disengagement of attention from the road}. In \bibinfo{booktitle}{\emph{Virtual, Augmented and Mixed Reality. Systems and Applications: 5th International Conference, VAMR 2013, Held as Part of HCI International 2013, Las Vegas, NV, USA, July 21-26, 2013, Proceedings, Part II 5}}. Springer, \bibinfo{pages}{261--270}.
\newblock


\bibitem[Olson et~al\mbox{.}(2009)]%
        {olson2009driver}
\bibfield{author}{\bibinfo{person}{Rebecca~L Olson}, \bibinfo{person}{Richard~J Hanowski}, \bibinfo{person}{Jeffrey~S Hickman}, \bibinfo{person}{Joseph Bocanegra}, {et~al\mbox{.}}} \bibinfo{year}{2009}\natexlab{}.
\newblock \bibinfo{booktitle}{\emph{Driver distraction in commercial vehicle operations}}.
\newblock \bibinfo{type}{{T}echnical {R}eport}. \bibinfo{institution}{United States. Department of Transportation. Federal Motor Carrier Safety~…}.
\newblock


\bibitem[Riegler et~al\mbox{.}(2021)]%
        {riegler2021augmented}
\bibfield{author}{\bibinfo{person}{Andreas Riegler}, \bibinfo{person}{Andreas Riener}, {and} \bibinfo{person}{Clemens Holzmann}.} \bibinfo{year}{2021}\natexlab{}.
\newblock \showarticletitle{Augmented reality for future mobility: Insights from a literature review and hci workshop}.
\newblock \bibinfo{journal}{\emph{i-com}} \bibinfo{volume}{20}, \bibinfo{number}{3} (\bibinfo{year}{2021}), \bibinfo{pages}{295--318}.
\newblock


\bibitem[Rzayev et~al\mbox{.}(2018)]%
        {rzayev2018reading}
\bibfield{author}{\bibinfo{person}{Rufat Rzayev}, \bibinfo{person}{Pawe{\l}~W Wo{\'z}niak}, \bibinfo{person}{Tilman Dingler}, {and} \bibinfo{person}{Niels Henze}.} \bibinfo{year}{2018}\natexlab{}.
\newblock \showarticletitle{Reading on smart glasses: The effect of text position, presentation type and walking}. In \bibinfo{booktitle}{\emph{Proceedings of the 2018 CHI conference on human factors in computing systems}}. \bibinfo{pages}{1--9}.
\newblock


\bibitem[Sawyer et~al\mbox{.}(2014)]%
        {sawyer2014google}
\bibfield{author}{\bibinfo{person}{Ben~D Sawyer}, \bibinfo{person}{Victor~S Finomore}, \bibinfo{person}{Andres~A Calvo}, {and} \bibinfo{person}{Peter~A Hancock}.} \bibinfo{year}{2014}\natexlab{}.
\newblock \showarticletitle{Google glass: A driver distraction cause or cure?}
\newblock \bibinfo{journal}{\emph{Human factors}} \bibinfo{volume}{56}, \bibinfo{number}{7} (\bibinfo{year}{2014}), \bibinfo{pages}{1307--1321}.
\newblock


\bibitem[Shahriar and Kun(2018)]%
        {shahriar2018camera}
\bibfield{author}{\bibinfo{person}{S~Tarek Shahriar} {and} \bibinfo{person}{Andrew~L Kun}.} \bibinfo{year}{2018}\natexlab{}.
\newblock \showarticletitle{Camera-view augmented reality: Overlaying navigation instructions on a real-time view of the road}. In \bibinfo{booktitle}{\emph{Proceedings of the 10th International Conference on Automotive User Interfaces and Interactive Vehicular Applications}}. \bibinfo{pages}{146--154}.
\newblock


\bibitem[Stahl(1999)]%
        {stahl1999amplitude}
\bibfield{author}{\bibinfo{person}{John~S Stahl}.} \bibinfo{year}{1999}\natexlab{}.
\newblock \showarticletitle{Amplitude of human head movements associated with horizontal saccades}.
\newblock \bibinfo{journal}{\emph{Experimental brain research}}  \bibinfo{volume}{126} (\bibinfo{year}{1999}), \bibinfo{pages}{41--54}.
\newblock


\bibitem[Topliss et~al\mbox{.}(2018)]%
        {topliss2018establishing}
\bibfield{author}{\bibinfo{person}{Bethan~Hannah Topliss}, \bibinfo{person}{Sanna~M Pampel}, \bibinfo{person}{Gary Burnett}, \bibinfo{person}{Lee Skrypchuk}, {and} \bibinfo{person}{Chrisminder Hare}.} \bibinfo{year}{2018}\natexlab{}.
\newblock \showarticletitle{Establishing the role of a virtual lead vehicle as a novel augmented reality navigational aid}. In \bibinfo{booktitle}{\emph{Proceedings of the 10th International Conference on Automotive User Interfaces and Interactive Vehicular Applications}}. \bibinfo{pages}{137--145}.
\newblock


\bibitem[van Amersfoorth et~al\mbox{.}(2019)]%
        {10.1145/3349263.3351911}
\bibfield{author}{\bibinfo{person}{Emma van Amersfoorth}, \bibinfo{person}{Lotte Roefs}, \bibinfo{person}{Quinta Bonekamp}, \bibinfo{person}{Laurent Schuermans}, {and} \bibinfo{person}{Bastian Pfleging}.} \bibinfo{year}{2019}\natexlab{}.
\newblock \showarticletitle{Increasing driver awareness through translucency on windshield displays}. In \bibinfo{booktitle}{\emph{Proceedings of the 11th International Conference on Automotive User Interfaces and Interactive Vehicular Applications: Adjunct Proceedings}} (Utrecht, Netherlands) \emph{(\bibinfo{series}{AutomotiveUI '19})}. \bibinfo{publisher}{Association for Computing Machinery}, \bibinfo{address}{New York, NY, USA}, \bibinfo{pages}{156–160}.
\newblock
\showISBNx{9781450369206}
\urldef\tempurl%
\url{https://doi.org/10.1145/3349263.3351911}
\showDOI{\tempurl}


\bibitem[Vogel et~al\mbox{.}(2009)]%
        {vogel2009hypoled}
\bibfield{author}{\bibinfo{person}{Uwe Vogel}, \bibinfo{person}{Ian Underwood}, \bibinfo{person}{Gunther Notni}, \bibinfo{person}{Christian Zilstorff}, \bibinfo{person}{Klaus Meerholz}, {and} \bibinfo{person}{Gunther Haas}.} \bibinfo{year}{2009}\natexlab{}.
\newblock \showarticletitle{HYPOLED-VGA OLED Microdisplay for HMD and Micro-projection}.
\newblock \bibinfo{journal}{\emph{IMID2009}} (\bibinfo{year}{2009}).
\newblock


\bibitem[von Sawitzky et~al\mbox{.}(2019)]%
        {10.1145/3321335.3324947}
\bibfield{author}{\bibinfo{person}{Tamara von Sawitzky}, \bibinfo{person}{Philipp Wintersberger}, \bibinfo{person}{Andreas Riener}, {and} \bibinfo{person}{Joseph~L. Gabbard}.} \bibinfo{year}{2019}\natexlab{}.
\newblock \showarticletitle{Increasing trust in fully automated driving: route indication on an augmented reality head-up display}. In \bibinfo{booktitle}{\emph{Proceedings of the 8th ACM International Symposium on Pervasive Displays}} (Palermo, Italy) \emph{(\bibinfo{series}{PerDis '19})}. \bibinfo{publisher}{Association for Computing Machinery}, \bibinfo{address}{New York, NY, USA}, Article \bibinfo{articleno}{6}, \bibinfo{numpages}{7}~pages.
\newblock
\showISBNx{9781450367516}
\urldef\tempurl%
\url{https://doi.org/10.1145/3321335.3324947}
\showDOI{\tempurl}


\bibitem[Wang et~al\mbox{.}(2023)]%
        {10.1145/3581961.3609870}
\bibfield{author}{\bibinfo{person}{Chao Wang}, \bibinfo{person}{Derck Chu}, {and} \bibinfo{person}{Marieke Martens}.} \bibinfo{year}{2023}\natexlab{}.
\newblock \showarticletitle{Enhancing Perception of Risk Objects for Car Drivers Through Augmented Reality Glasses}. In \bibinfo{booktitle}{\emph{Adjunct Proceedings of the 15th International Conference on Automotive User Interfaces and Interactive Vehicular Applications}} (Ingolstadt, Germany) \emph{(\bibinfo{series}{AutomotiveUI '23 Adjunct})}. \bibinfo{publisher}{Association for Computing Machinery}, \bibinfo{address}{New York, NY, USA}, \bibinfo{pages}{83–86}.
\newblock
\showISBNx{9798400701122}
\urldef\tempurl%
\url{https://doi.org/10.1145/3581961.3609870}
\showDOI{\tempurl}


\bibitem[Yu et~al\mbox{.}(2022)]%
        {yu2022gallium}
\bibfield{author}{\bibinfo{person}{Junchi Yu}, \bibinfo{person}{Feifan Xu}, \bibinfo{person}{Tao Tao}, \bibinfo{person}{Bin Liu}, \bibinfo{person}{Bin Wang}, \bibinfo{person}{Yimeng Sang}, \bibinfo{person}{Shihao Liang}, \bibinfo{person}{Yang Chen}, \bibinfo{person}{Meixin Feng}, \bibinfo{person}{Zhe Zhuang}, {et~al\mbox{.}}} \bibinfo{year}{2022}\natexlab{}.
\newblock \showarticletitle{Gallium nitride blue/green micro-LEDs for high brightness and transparency display}.
\newblock \bibinfo{journal}{\emph{IEEE Electron Device Letters}} \bibinfo{volume}{44}, \bibinfo{number}{2} (\bibinfo{year}{2022}), \bibinfo{pages}{281--284}.
\newblock


\end{thebibliography}

\end{document}